\documentclass[twocolumn,showpacs,preprintnumbers,amssymb,amsmath,superscriptaddress,letterpaper,nofootinbib]{revtex4-1}[10pt]

\usepackage{graphicx}
\usepackage{dcolumn}
\usepackage{bm}
\usepackage{pstricks}
\usepackage{color}
\usepackage{epsfig}

\newcommand{\ba}{\begin{array}}
\newcommand{\ea}{\end{array}}
\newcommand{\bqa}{\begin{eqnarray}}
\newcommand{\eqa}{\end{eqnarray}}

\newcommand{\be}{\begin{equation}}
\newcommand{\ee}{\end{equation}}
\newcommand{\bea}{\begin{eqnarray}}
\newcommand{\eea}{\end{eqnarray}}

\def\draft{
}

\begin{document}
\draft

\title{Testing The Light Dark Matter Hypothesis With AMS}

\author{Dan Hooper}
%\email{dhooper@fnal.gov}
\affiliation{Theoretical Astrophysics Group, Fermi National Accelerator Laboratory, Batavia, IL  60510}
\affiliation{Department of Astronomy and Astrophysics, University of Chicago, Chicago, IL 60637}

\author{Wei Xue}
\affiliation{Department of Physics, McGill University, 3600 Rue University,  Montr\'eal, Qu\'ebec, Canada H3A 2T8}

\date{\today}

\begin{abstract}

The spectrum and morphology of gamma-rays from the Galactic Center and the spectrum of synchrotron emission observed from the Milky Way's radio filaments have each been interpreted as possible signals of $\sim$7-10 GeV dark matter particles annihilating in the Inner Galaxy. In dark matter models capable of producing these signals, the annihilations should also generate significant fluxes of $\sim$7-10 GeV positrons which can lead to a distinctive bump-like feature in local cosmic ray positron spectrum. In this letter, we show that while such a feature would be difficult to detect with PAMELA, it would likely be identifiable by the currently operating AMS experiment. As no known astrophysical sources or mechanisms are likely to produce such a sharp feature, the observation of a positron bump at around 7-10 GeV would significantly strengthen the case for a dark matter interpretation of the reported gamma-ray and radio anomalies.

\end{abstract}
\pacs{95.35.+d, 98.70.Sa, 96.50.S\hfill FERMILAB-PUB-12-534-A}
\maketitle

%\section{Introduction}

Dark matter particles annihilating in the halo of the Milky Way can potentially lead to enhanced quantities of antimatter in the cosmic ray spectrum. PAMELA's observation of a rising cosmic ray positron fraction (defined as the ratio of positrons-to-positrons plus electrons) between $\sim$10-100 GeV~\cite{Adriani:2008zr} generated a great deal of excitement focused around this possibility. In the light of more recent measurements of the cosmic ray electron spectrum from Fermi and HESS~\cite{Abdo:2009zk,Aharonian:2009ah}, as well as constraints such as those from cosmic ray antiprotons~\cite{Adriani:2008zq}, however, it now appears more likely that the observed rise is the product of astrophysical phenomena, such as nearby pulsars~\cite{Hooper:2008kg,Profumo:2008ms}.

\begin{figure*}[!]
\vspace{-2.8cm}
\includegraphics[width=0.36\textwidth ]{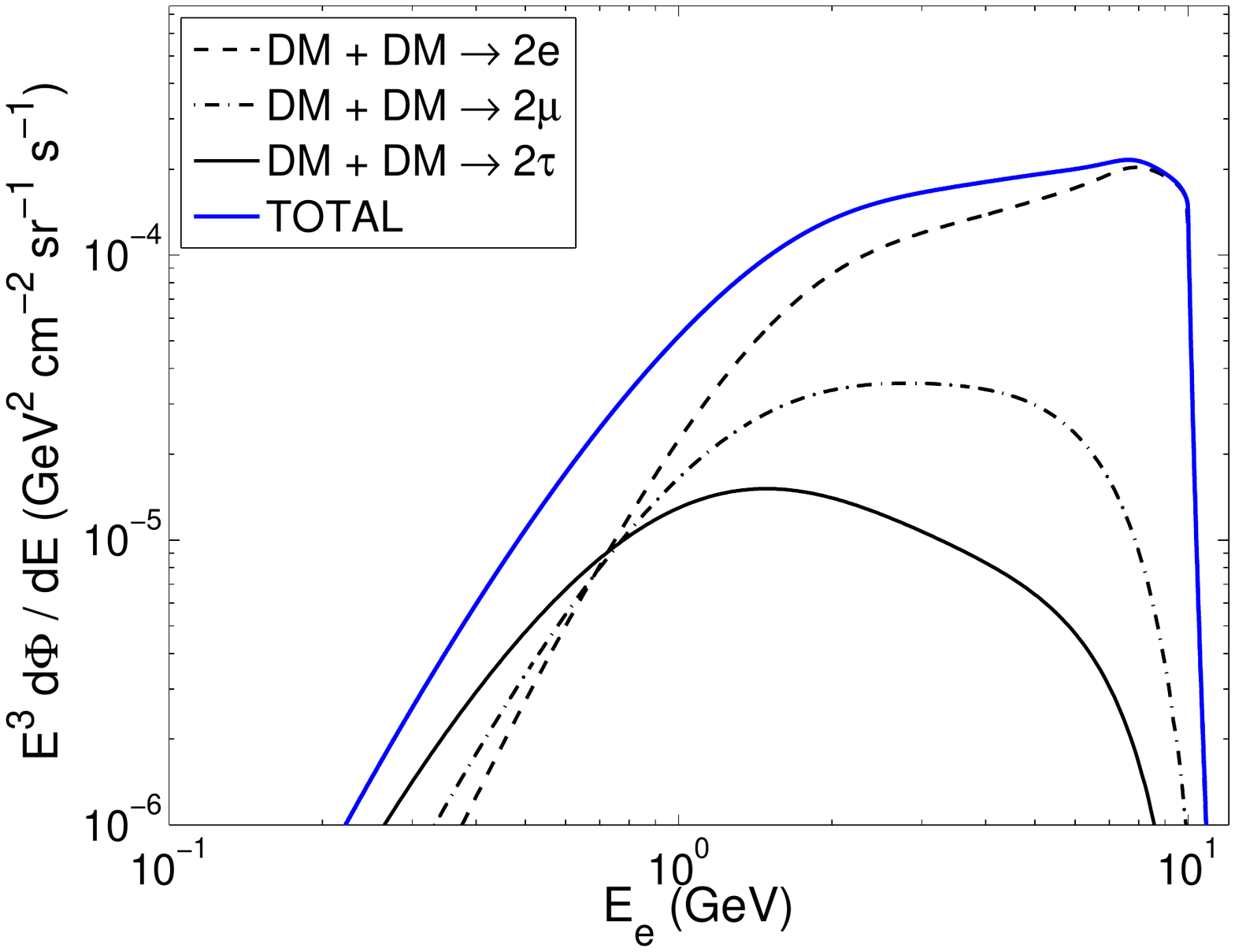}
\hspace{-1.0cm}
\includegraphics[width=0.36\textwidth ]{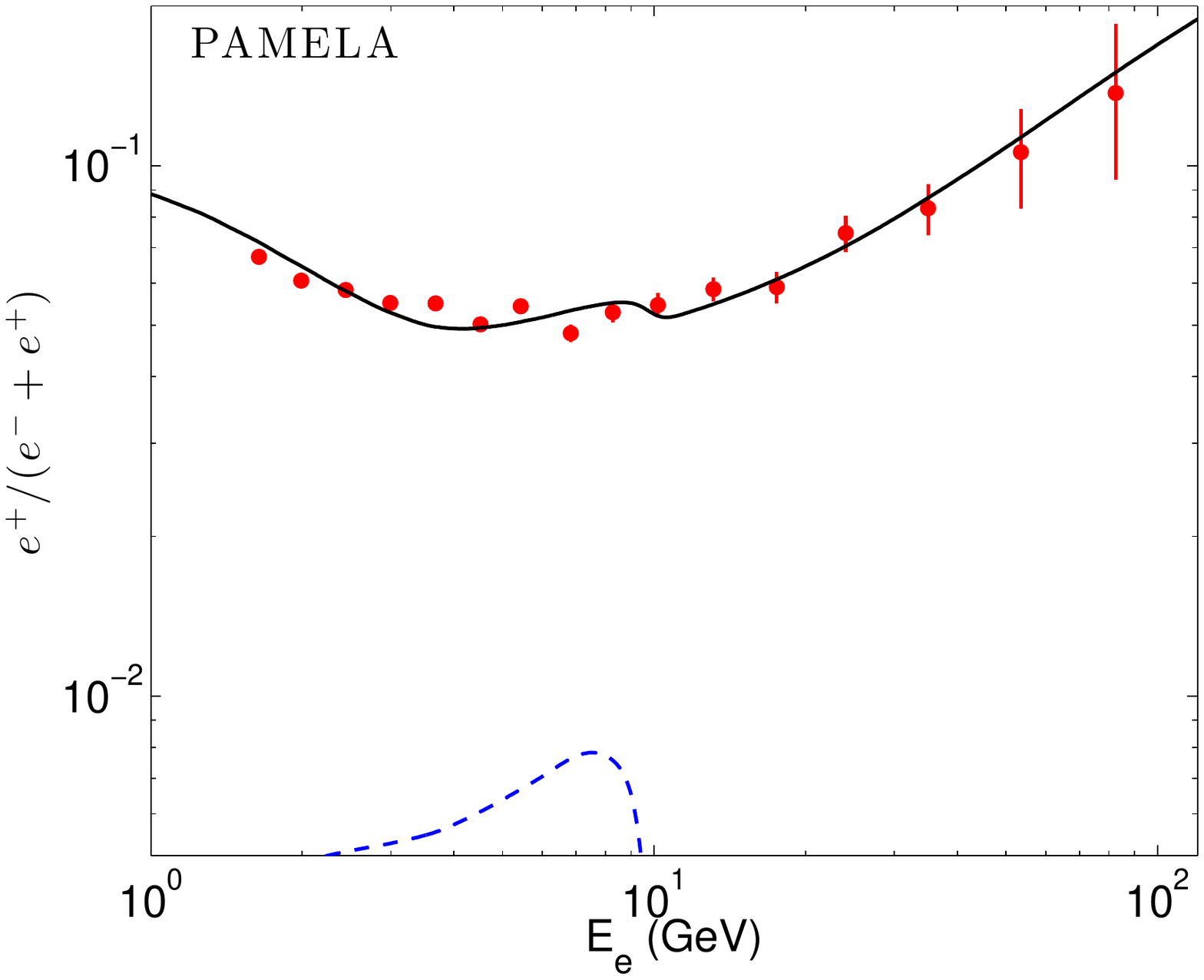} 
\hspace{-1.0cm}
\includegraphics[width=0.36\textwidth ]{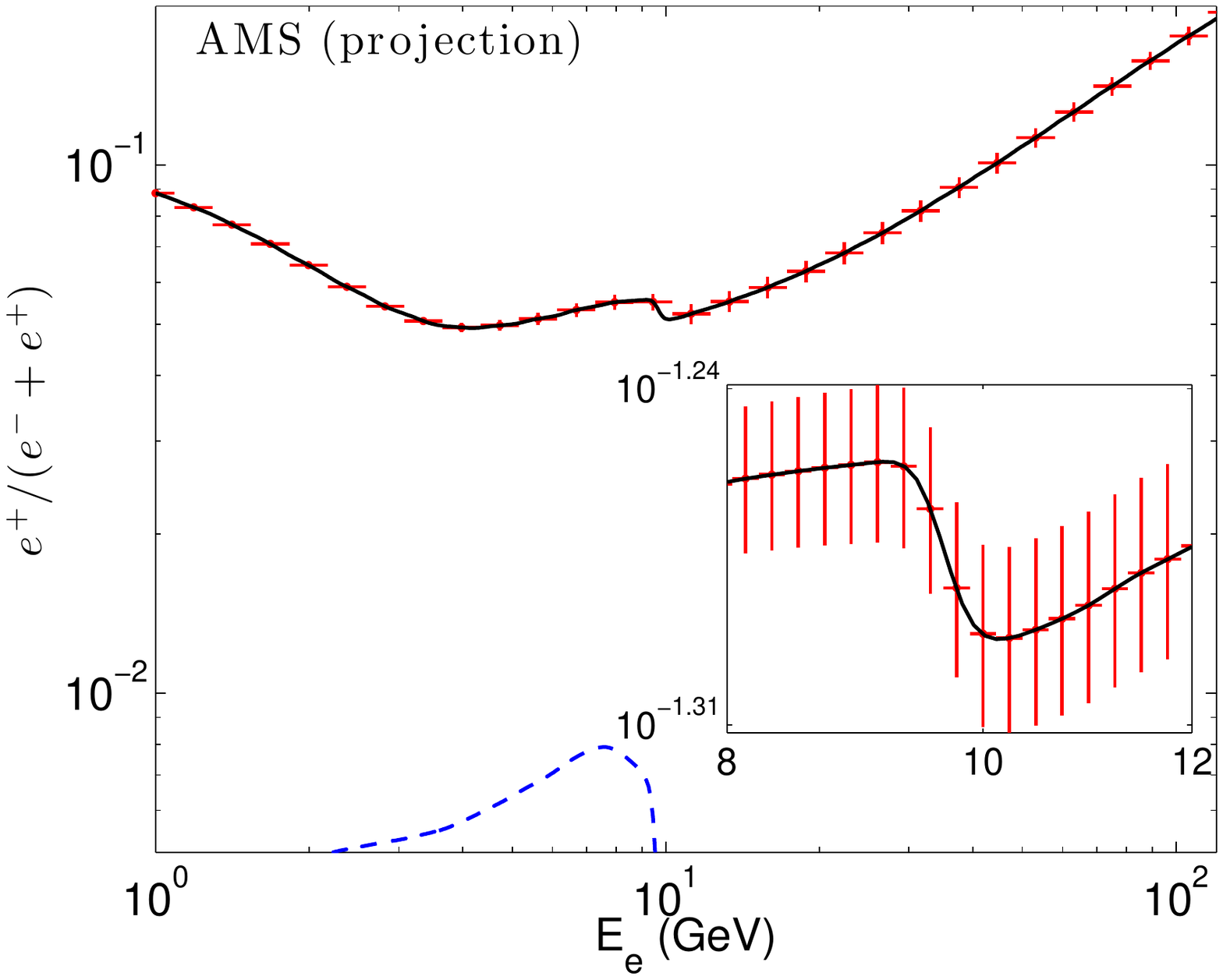} 
\vspace{-2.2cm}
\caption{Left: The contribution to the local cosmic ray positron spectrum from 10 GeV dark matter particles annihilating democratically to charged lepton pairs, neglecting the effects of solar modulation. Center: The cosmic ray positron fraction predicted in this model compared to the measurements of PAMELA, including astrophysical backgrounds from secondary production and a nearby pulsar, and including the effects of solar modulation. The dashed blue line denotes the contribution from dark matter annihilations. Right: The projected ability of AMS to measure the cosmic ray positron fraction in this scenario. The distinctive feature at an energy equal to the dark matter mass can clearly be identified by AMS. In each frame, we have adopted a generalized NFW profile with an inner slope of $\rho_{\rm DM} \propto r^{-1.2}$  and an annihilation cross section chosen to match the gamma-ray and radio signals observed from the inner Galaxy ($\sigma v= 4.5\times 10^{-27}$ cm$^3$/s). See text for more details.}
\label{leptonflux}
\end{figure*}

We are eagerly awaiting the first science results of the Alpha Magnetic Spectrometer (AMS-02) experiment. AMS will measure with unprecedented precision many of the components of the cosmic ray spectrum between approximately 100 MeV and 1 TeV. In particular, with its much larger acceptance than PAMELA ($\sim$0.045 m$^2$sr vs. $\sim$0.002 m$^2$sr) and its high level of proton rejection, AMS is expected to measure the cosmic ray positron and antiproton spectra in far greater detail than was previously possible. Furthermore, by providing better measurements of various cosmic ray secondary-to-primary ratios (such as boron-to-carbon and beryllium-10-to-beryllium-9), AMS will be able to much more tightly constrain the underlying diffusion model~\cite{Pato:2010ih} and thus improve our ability to predict the spectra of cosmic ray antimatter that may result from dark matter annihilations in the galactic halo.

Dark matter particles which annihilate mostly to quarks or gauge bosons yield a largely featureless spectrum of positrons which, after accounting for the effects of cosmic ray propagation, are likely to be difficult to separate from other sources of cosmic ray positrons. This is exacerbated by the large flux of (likely astrophysical) high energy positrons observed by PAMELA (and subsequently by Fermi~\cite{FermiLAT:2011ab}). A different conclusion can be reached, however, in cases in which the dark matter particles annihilate directly to electron-positron pairs, leading to a sudden edge-like feature in the cosmic ray positron spectrum at an energy equal to the mass of the annihilating WIMP~\cite{Baltz:2004ie}. Such a feature could plausibly go unidentified by PAMELA, while being readily detectable with AMS.

Recently, an intriguing body of evidence has accumulated in favor of relatively light dark matter particles which annihilate largely to leptons, including to $e^+ e^-$~\cite{Hooper:2012ft}. In particular, the spectrum and angular distribution of gamma-rays observed from the region surrounding the Galactic Center can be well fit by a 7-10 GeV dark matter particle, distributed with a cusped halo profile, and annihilating to leptons with a cross section on the order of $\sigma v \sim 10^{-26}$ cm$^3$/s~\cite{Hooper:2011ti,Hooper:2010mq,Abazajian:2012mx}. The same dark matter model (particle and distribution) can also naturally explain the peculiar radio emission observed from the Milky Way's radio filaments~\cite{Linden:2011au}, and could account for much or most of the Milky Way's synchrotron haze~\cite{Hooper:2010im}.\footnote{Direct detection anomalies reported by DAMA/LIBRA, CoGeNT, and CRESST may also be able to be explained by the same 7-10 GeV dark matter candidate, although this has little bearing on the results of this study (see Ref.~\cite{Kelso:2011gd} and references therein).} In this letter, we show that in scenarios in which dark matter explains these signals, one expects AMS to observe a distinctive feature at around 7-10 GeV in the cosmic ray positron spectrum (and positron fraction), although such a feature would likely be unresolvable by PAMELA. 

We begin by considering a simple phenomenological dark matter model that is capable of explaining the aforementioned gamma-ray and radio signals. In particular, we consider a model in which the dark matter consists of a 7-10 GeV particle which annihilate democratically to charged lepton pairs. The decays of the tau-leptons produce a gamma-ray spectrum consistent with that observed from the Galactic Center~\cite{Hooper:2011ti}, while the electrons and positrons generate the synchrotron emission from the observed radio filaments~\citep{Linden:2011au}. For possible realizations of such phenomenological features within a particle physics model, see Ref.~\cite{Buckley:2010ve}. To accommodate the observed morphology of gamma-ray and synchrotron emission from the Inner Galaxy, we adopt a dark matter distribution which follows a generalized NFW profile with an inner slope of $\rho_{\rm DM} \propto r^{-1.2}$ and a scale radius of 20 kpc. To normalize these signals, we adopt an annihilation cross section of $\sigma v= 4.5 \times 10^{-27}$ cm$^3$/s and a local dark matter density of 0.4 GeV/cm$^3$.

Once injected into the halo, electrons and positrons diffuse through the Galactic Magnetic Field, steadily losing energy through a combination of inverse Compton scattering and synchrotron losses. To determine the cosmic ray spectrum as observed at the Solar System, we solve the standard propagation equation (using the publicly available code GALPROP):
\bqa 
\frac{\partial   \psi }{\partial	 t } &=& Q(\textbf{r},p) + \bigtriangledown \cdot \left( D_{xx} \bigtriangledown \psi  - \textbf{V} \psi \right) +
\frac{\partial }{ \partial p  } p^2 D_{pp} \frac{\partial }{ \partial p  } \frac{1}{p^2} \psi   \nonumber\\
&& - \frac{\partial }{ \partial p  } \left[\dot{p} \psi -\frac{p}{3} \left( \bigtriangledown \cdot \textbf{V} \right) \psi \right]  
- \frac{1}{\tau_f} -\frac{1}{\tau_r} \psi   \ ,
\label{propeq}
\eqa
where $\psi(\textbf{r},p,t)$ is the number density of a given cosmic ray species per unit momentum, and the source term $Q(\textbf{r},p)$ includes the products of the decay and spallation of nuclei, as well as any primary contributions from supernova remnants, pulsars, dark matter annihilations, etc. $D_{xx}$ is the spatial diffusion coefficient, which is parametrized by $D_{xx} = \beta D_{0xx} (\rho/4 GV)^\delta$, where $\beta$ and $\rho$ are the particle's velocity and rigidity, respectively. Also included in this equation are the effects of diffusive reacceleration, convection, and radioactive decay~\cite{Strong:1998pw}. The contribution to the source term, $Q(\textbf{r},p)$, from dark matter is simply determined by the flux of annihilation products injected into the halo. In our calculations, we adopt $D_{0xx}=5.25 \times 10^{28}$ cm$^2$/s and apply free-escape boundary conditions at 4 kpc above and below the Galactic Plane. These choices lead to boron-to-carbon and antiproton-to-proton ratios that are consistent with observations.

For the electron/positron energy loss rate, we include contributions from the default GALPROP radiation field model, and from a magnetic field model described by $B = 7 \mu{\rm G} \,  \exp(-r/10 \, \mathrm{kpc}) \exp(-|z|/2 \, \mathrm{kpc})$, where $r$ and $z$ describe the location in galactic (cylindrical) coordinates.

%%%

%%%

\begin{figure*}[!]
\vspace{-2.8cm}
\includegraphics[width=0.36\textwidth ]{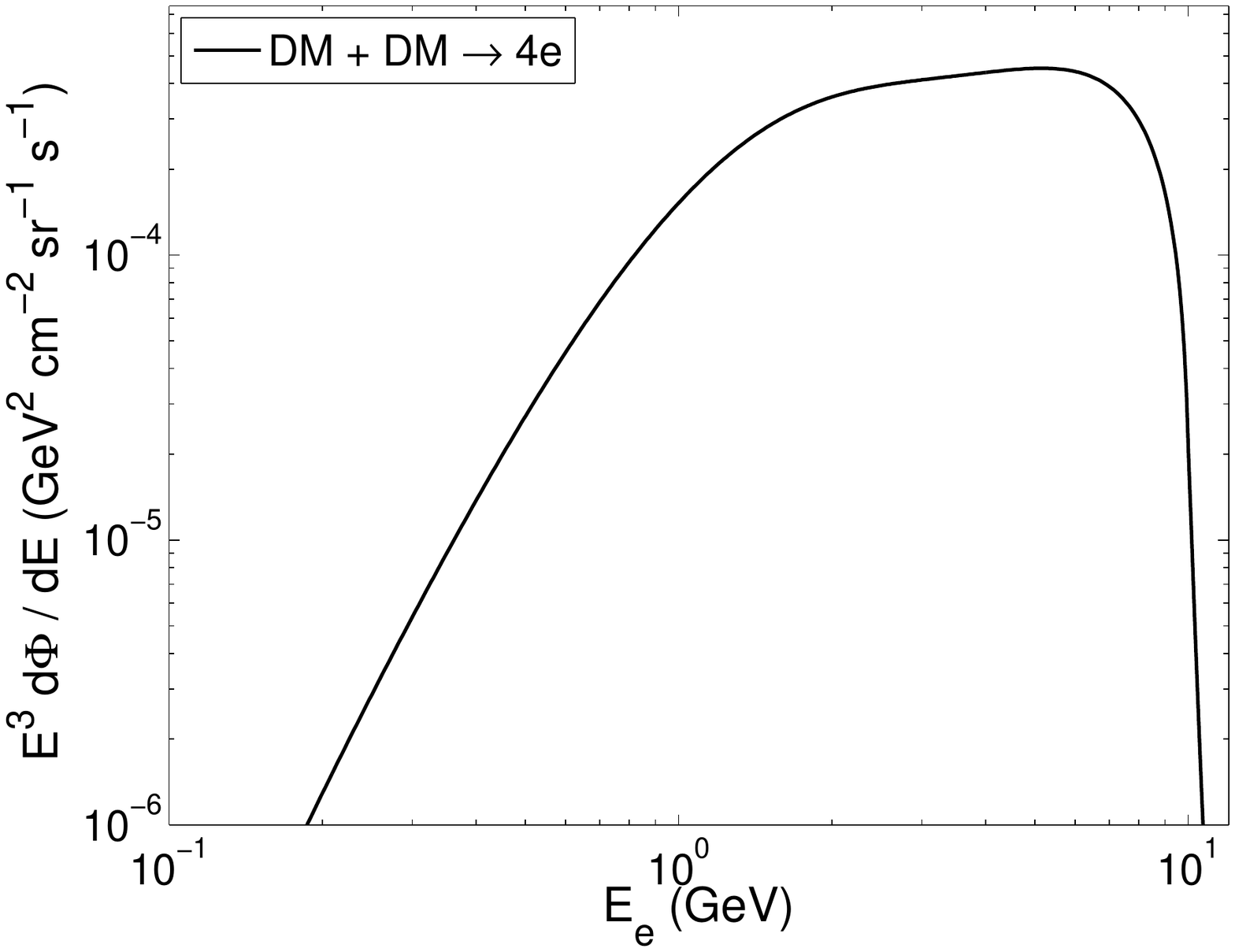}
\hspace{-1.0cm}
\includegraphics[width=0.36\textwidth ]{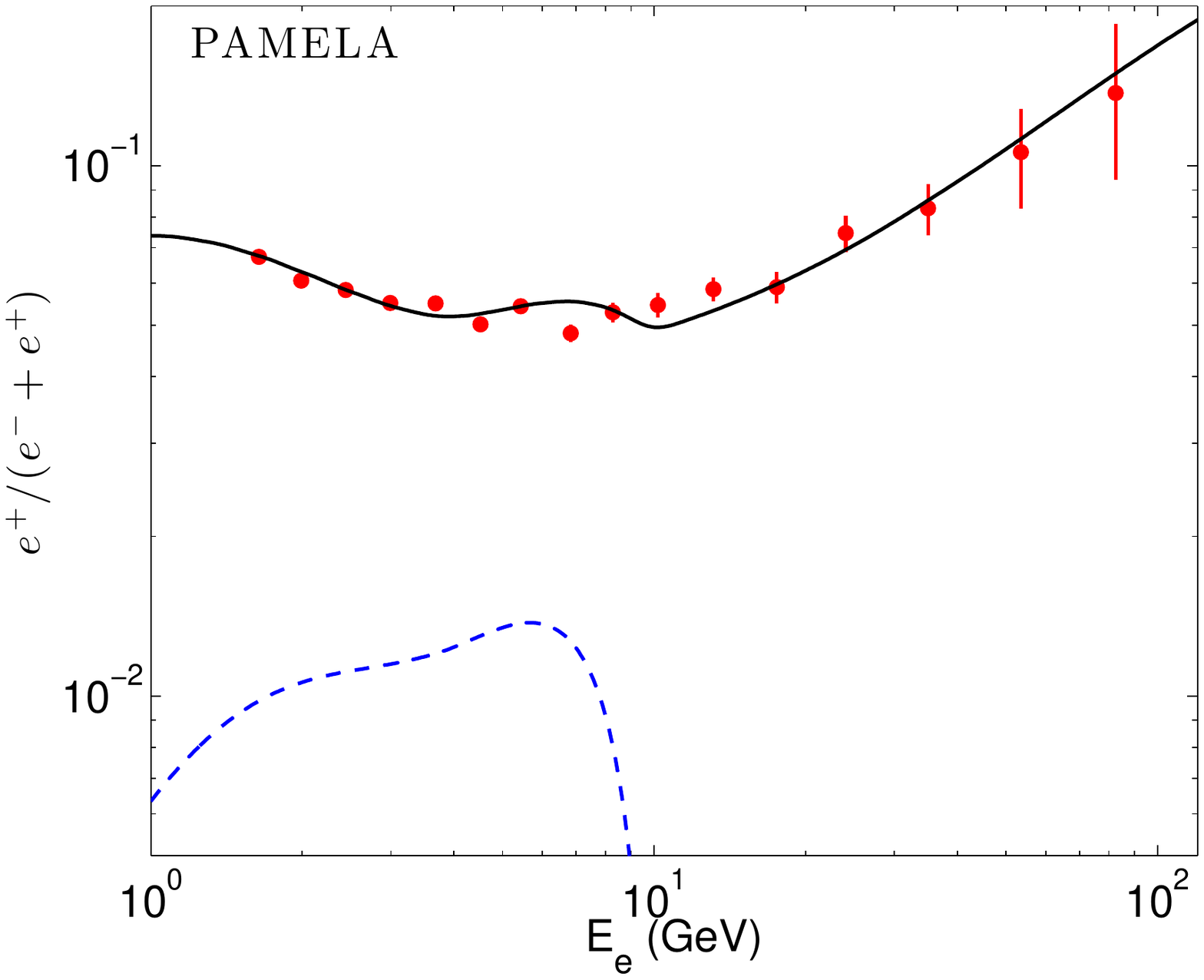}
\hspace{-1.0cm}
\includegraphics[width=0.36\textwidth ]{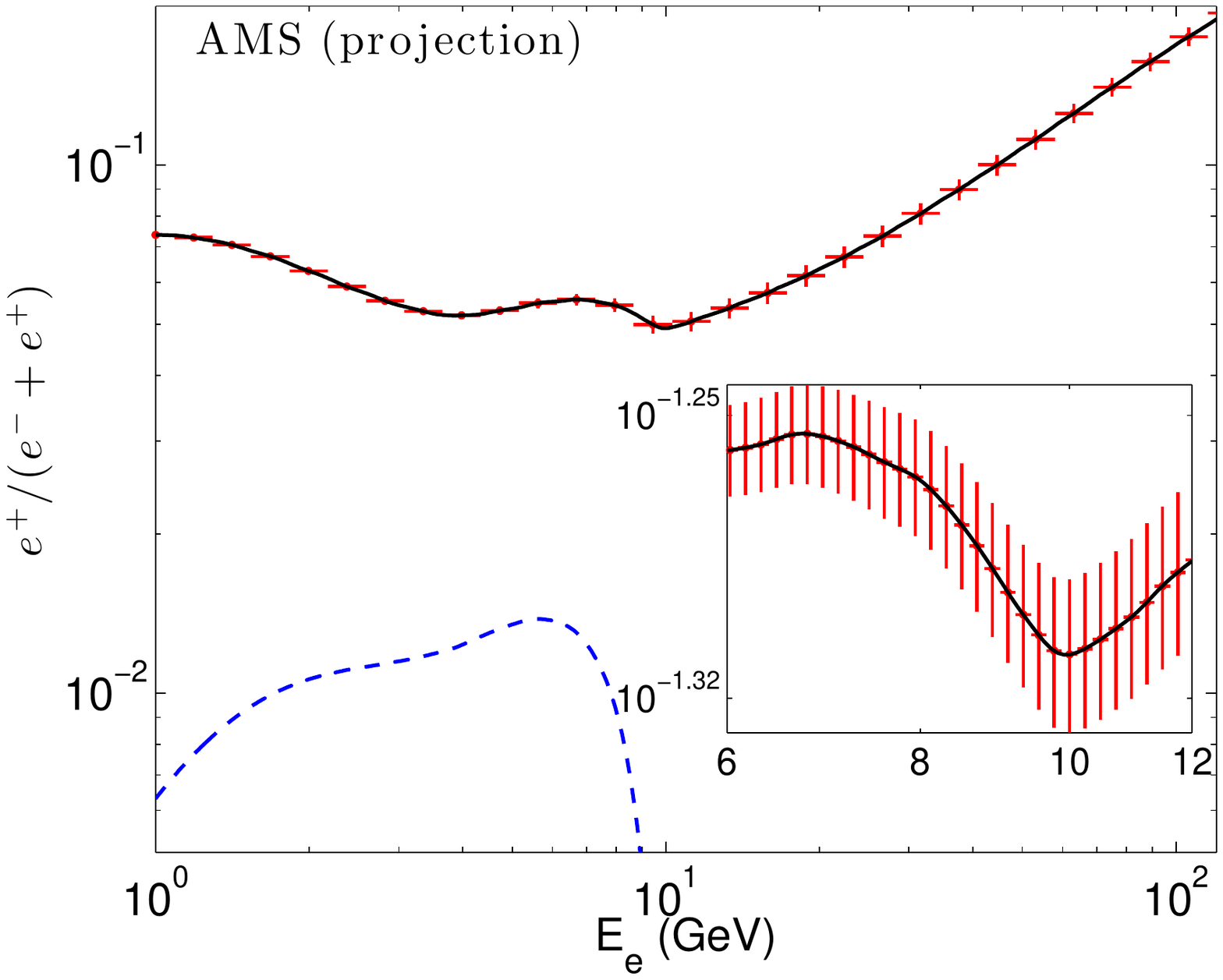}
\vspace{-2.2cm}
\caption{As in Fig.~\ref{leptonflux}, for the case of dark matter particles which annihilate to a pair of 200 MeV gauge bosons (which then decay to electron-positron pairs). Again, the annihilation cross section has been chosen to match the gamma-ray and radio signals observed from the inner Galaxy ($\sigma v= 6.6\times 10^{-27}$ cm$^3$/s).}
\label{DF}
\end{figure*}

In the left frame of Fig.~\ref{leptonflux}, we show the contribution to the local cosmic ray positron spectrum from dark matter annihilations. Note the sudden drop in the cosmic ray positron flux at 10 GeV (corresponding to the mass of the dark matter particle). The dark matter contribution to the flux of positrons at energies just below the edge can be calculated analytically and is given by~\cite{Baltz:2004ie}:
\begin{equation}
\frac{d\Phi_{e^+}}{dE_{e^+}}\bigg|_{\rm edge} =  \frac{c}{8\pi}\frac{\sigma_{e^+e^-} v}{(dE_e/dt)} \bigg(\frac{\rho_{\rm DM}}{m_{\rm DM}}\bigg)^2,
\end{equation}
where $\sigma_{e^+e^-} v$ is the dark matter annihilation cross section to electron-positron pairs, $\rho_{\rm DM}$ is the local density of dark matter, and $dE_e/dt$ is the local energy loss rate of electrons/positrons from synchrotron and inverse Compton scattering. This energy loss rate can be written in terms of the local densities of radiation and magnetic fields:
\begin{equation}
\frac{dE_e}{dt} \approx 1.02 \times 10^{-14} \, {\rm GeV/s} \,  \bigg(\frac{\rho_{\rm rad}+\rho_{B}}{1\,{\rm eV/cm}^3}\bigg) \, \bigg(\frac{E_e}{10 \, {\rm GeV}}\bigg)^2.
\end{equation}
Combining these two equations, and for a local energy density in radiation and magnetic fields of 1.4 eV/cm$^3$, this predicts a positron flux at the dark matter's mass of $d\Phi_{e^+}/dE_{e^+} \approx 2\times 10^{-7}$ cm$^{-2}$ sr$^{-1}$ s$^{-1}$ GeV$^{-1}$. As we will demonstrate, this sudden drop will lead to a distinctive feature in the positron fraction, likely observable to AMS.

To evaluate the prospects for observing such a contribution to the cosmic ray positron spectrum, we must consider the relevant astrophysical backgrounds, as well as the effects of solar modulation. In our analysis, we will adopt a simple background model which includes contributions to the cosmic ray positron spectrum from both secondary production ({\it ie}. positrons from the interactions of cosmic ray protons) and from a nearby pulsar. In particular, we consider the Geminga pulsar which is located 157 parsecs from the Solar System and is 370,000 years old. We assume this pulsar to have injected a spectrum of positrons and electrons of the form, $Q  \propto E_e^{-1.55}$. We follow Ref.~\cite{Hooper:2008kg} in determining the flux of positrons and electrons at the Solar System from this pulsar and add this to the contribution predicted from secondary production, as obtained using GALPROP.

As cosmic rays enter the Solar System, they are further impacted by solar winds and the helioshperic magnetic field. These effects can be modeled by a charge-sign dependent effective potential, $\Phi = (eZ/A) \phi_{\pm}$, where $Z$ and $A$ are the charge and atomic number of the cosmic ray species~\cite{1968ApJ...154.1011G,Moskalenko:2001ya,Beischer:2009zz}. We adopt a modulation potential described by $\phi_{+}=400$ MV and $\phi_{-}=180$ MV. Note that while the effective modulation potential is expected to vary with time within the solar cycle, we use the same values of $\phi_{\pm}$ for AMS projections as we did for the PAMELA case. As such variations will only smoothly alter the cosmic ray positron spectrum and will not induce any distinctive features, this simplifying assumption should not significantly impact our conclusions.

In the center frame of Fig.~\ref{leptonflux}, we show the resulting positron fraction and compare this to that measured by PAMELA. Note that although the dark matter annihilation products do lead to a bump in the positron spectrum at about the mass of the WIMP, this feature is modest enough to likely go unidentified by PAMELA. In the right frame of the same figure, we project the ability of AMS to measure such a feature in the cosmic ray positron fraction. To project the error bars for AMS, we follow Ref.~\cite{Pato:2010im}. In particular, we convolve the spectrum of the positron fraction with an energy resolution of $\Delta E/E = \sqrt{(0.106/\sqrt{E({\rm GeV})})^2+(0.0125)^2}$ (corresponding to about 3.5\% at 10 GeV), an ability to reject protons from positrons at the level of $3 \times 10^5$~\cite{Casaus:2009zz}, and an acceptance of 0.045 m$^2$ sr. And while we have calculated our error bars for 1000 days of data taking, the systematic rather than statistical errors dominate the results.

In the insert of the right frame of Fig.~\ref{leptonflux}, we focus in on the energy range around the feature, and bin the projected data more finely in this region. We find that AMS should be able to clearly identify the presence of such a bump-like feature. Interestingly, unlike more smoothly varying contributions to the positron fraction, it would be challenging to attribute such a sudden change in the positron fraction to astrophysical sources or mechanisms. Instead, the observation of such a sudden drop would require the injection of a nearly mono-energetic spectrum of positrons in the local region of our galaxy. And while this is very unexpected from astrophysical sources, dark matter annihilating directly to electrons and positrons would be generically expected to produce such a spectral feature.

%%%%%%%%%%%%%%%%%%%%%%%%%%%%%%%%%%%%%

So far, we have considered dark matter candidates which annihilate directly to electron positron pairs a significant fraction of the time. And while we have demonstrated that such models lead to a distinctive feature that would likely be observable to AMS, we can also consider other dark matter models capable of producing the observed gamma-ray and synchrotron signals from the inner Galaxy that may be more difficult to observe with AMS. More specifically, we will also consider a model in which the dark matter annihilates to a pair of light gauge bosons, $\phi$, which then decay to Standard Model leptons and mesons through a small degree of kinetic mixing with the photon~\cite{Czyz:2008kw,Achasov:2002ud,Kambor:1991ah}. Such a scenario, which we will refer to as the dark forces model, was shown in Ref.~\cite{Hooper:2012cw} to be able to produce the observed gamma-ray and synchrotron signals. For concreteness, we will consider a value of $m_{\phi}=0.2$ GeV. For this choice, the gamma-rays from final state radiation lead to a signal compatible with that observed from the Galactic Center, while the decays to $e^+ e^-$ provide the synchrotron signal from the radio filaments. 

In Fig.~\ref{DF}, we show the resulting cosmic ray positron spectrum and positron fraction that results in this dark forces model. Here we have used slightly different solar modulation parameters ($\phi_+=310$ MV, $\phi_-=20$ MV) in order to obtain a good fit to the positron fraction observed by PAMELA. In this case, the four body final state that results from ${\rm DM}\,{\rm DM} \rightarrow \phi \, \phi \rightarrow e^+ e^- e^+ e^-$ leads to a somewhat softer spectral feature, without the sharp edge found in the previous case. Even in this more difficult model, however, the positron bump that is generated at $\sim$5-8 GeV should be discernible by AMS. We note, however, that if we had chosen a larger value of $m_{\phi}$, a smaller fraction of the total annihilation power would have gone into electrons/positrons, reducing the magnitude of this feature. For this reason, AMS will likely only be sensitive to the dark forces model for values of $m_{\phi}$ less than 1 GeV or so.

In summary, we have demonstrated that if dark matter annihilations are responsible for the anomalous gamma-rays observed from the Galactic Center and the synchrotron emission observed from the Milky Way's radio filaments, then dark matter annihilations taking place in the local halo should also produce a distinctive feature in the cosmic ray positron fraction at an energy of around 7-10 GeV. While such a feature would be difficult to resolve with PAMELA, the currently operating AMS experiment should be capable of clearly identifying it.

The presence of a sudden bump-like feature in the cosmic ray positron fraction would be difficult to attribute to astrophysical phenomena. In particular, in order to generate such a feature, a source of cosmic ray positrons would have to be located within a kiloparsec or less of the Solar System and inject an approximately mono-energetic spectrum of positrons. While this is not expected from any class of astrophysical sources or mechanisms, dark matter particles annihilating to electron-positron pairs are generically predicted to generate such a feature. If data from AMS reveals a feature in the cosmic ray positron fraction similar to that described in this letter, this would be most easily interpreted as evidence for dark matter particles annihilating to electrons/positrons, and would further strengthen the case for a dark matter interpretation of the gamma-ray and synchrotron signals observed from the Inner Galaxy.

%The conclusions reached here are not

%While a number of astrophysical uncertainties exist pertaining to the diffusion model and the effects of solar modulation, such considerations should only smoothly impact the positron spectrum, and should be capable of neither creating nor washing out the sudden and discrete spectral feature that is predicted in this scenario. The only astrophysics that conceal the presence of this feature would be if the local neighborhood contains a higher than expected density of radiation and/or magnetic fields, leading to a more rapid energy loss rate for positrons and suppressing the feature in question.

\vspace{1.0 cm}

%\vskip 0.2 in
%\noindent {\bf Acknowledgments}
\vskip 0.0001in
\noindent We would like to thank Miguel Pato, Gianfranco Bertone, Tim Linden, Ilias Cholis, and Aaron Vincent for helpful discussions. DH is supported by the US Department of Energy. WX is supported by the URA Visiting Scholars Program at Fermilab. This work was carried out in part at the Aspen Center for Physics, which is supported by the National Science Foundation under grant number PHY-1066293.

\newpage

 \bibliography{darkforce.bib}
 \bibliographystyle{h-physrev}

%%%%%%%%%%%

%\bibliography{pamela}
%\bibliographystyle{apsrev}

\end{document}